\documentclass[prb,preprintnumbers,preprint]{revtex4}
\usepackage[draft]{hyperref}
\usepackage{amsmath}
\usepackage{amssymb}
\usepackage{graphicx}

\begin{document}
\title{Isotropic and non-diffracting optical metamaterials}

\author{T. Paul, C. Menzel, C. Rockstuhl, and F.
Lederer} \affiliation{Institute of Condensed Matter Theory
and Optics, Friedrich-Schiller-Universit\"at Jena, Max-Wien-Platz 1,
D-07743 Jena, Germany}

\begin{abstract}
Optical metamaterials have the potential to control the flow
of light at will\cite{Metamaterials1,Metamaterials2,Metamaterials3}
which may lead to spectacular applications as the perfect
lens\cite{lens} or the cloaking device\cite{cloak}. Both of these
optical elements require invariant effective material properties
(permittivity, permeability) for all spatial frequencies involved in
the imaging process. However, it turned out that due to the
mesoscopic nature of current metamaterials spatial dispersion
prevents to meet this requirement\cite{Tretya}; rendering them far
away from being applicable for the purpose of
imaging\cite{dispersion,imaging}. A solution to this problem is not
straightforwardly at hand since metamaterials are usually designed
in forward direction; implying that the optical properties are only
evaluated for a specific metamaterial. Here we lift these
limitations. Methodically, we suggest a procedure to design
metamaterials with a predefined characteristic of light propagation.
Optically, we show that metamaterials can be optimized such that
they exhibit either an isotropic response or permit diffractionless
propagation.
\end{abstract}

\maketitle

Metamaterials are usually characterized as deriving their optical properties
predominantly from the geometry of subwavelength unit cells and attaining an
effective permittivity $\epsilon_{\textrm{eff}}(\omega)$ or permeability
$\mu_{\textrm{eff}}(\omega)$ not accessible by natural materials. In
particular, a sufficient condition for the existence of left-handed waves (wave
vector $\mathbf{k}$ is anti-parallel to the Poynting vector $\mathbf{S}$) as
eigenmodes is a negative real part of both effective parameters
\cite{lefthanded}. Then, at normal incidence an effective refractive index,
which is defined by normalizing the wavevector by $k_0=2\pi/\lambda$, may be
introduced and attains negative values, e.g. $n=-1$. Frequently, it is then
concluded that such a metamaterial might be employed as a perfect lens
\cite{pendry_lens}. But this conclusion implicitly assumes that this effective
index is invariant for all propagating and evanescent waves.

However, for all current optical metamaterials this requirement is
not fulfilled \cite{dispersion}. Most metamaterials perform worse
than an ordinary lens since even in the long wavelength limit their
refraction, diffraction and absorption properties are anisotropic.
Consequently, the effective index is far from being a suitable
quantity to describe propagation of light beams of finite spatial
width in metamaterials \cite{imaging}. One rather has to resort to
refraction and diffraction coefficients
\cite{refractiondiffraction}.

Consequently, a reasonable design approach towards metamaterials
should not rely on effective material parameters but on desired
light propagation characteristics or imaging properties not
available in nature. Thus a new design strategy must be developed
where a particular optical property of metamaterials represents the
essential target function.

Here we show that this design strategy can be based on the
dispersion relation of the fundamental Bloch wave which is an
eigenmode of a bulk metamaterial. Departing from its dispersion
relation $\omega  = \omega (k_x ,k_y ,k_z )$ one can impose desired
constraints such as an isotropic optical response (spherical
dispersion relation with constant negative curvature) or even a real
and imaginary part of the longitudinal wave vector component which
does not depend on the transverse one. The former property permits
the introduction of a constant effective negative index whereas the
latter one causes light to propagate diffractionless at a constant
attenuation, rendering the material amenable to a light tunneling
scheme for the purpose of imaging
\cite{lighttunneling1,lighttunneling2}. To achieve these design
goals the number of degrees of freedom defining a metamaterial has
to be increased. Numerical optimization techniques may be
subsequently applied to reveal an optimum set of parameters such
that the optical response of the metamaterial matches with
sufficient precision the predefined one.
\begin{figure}[h]
\includegraphics[width=8cm]{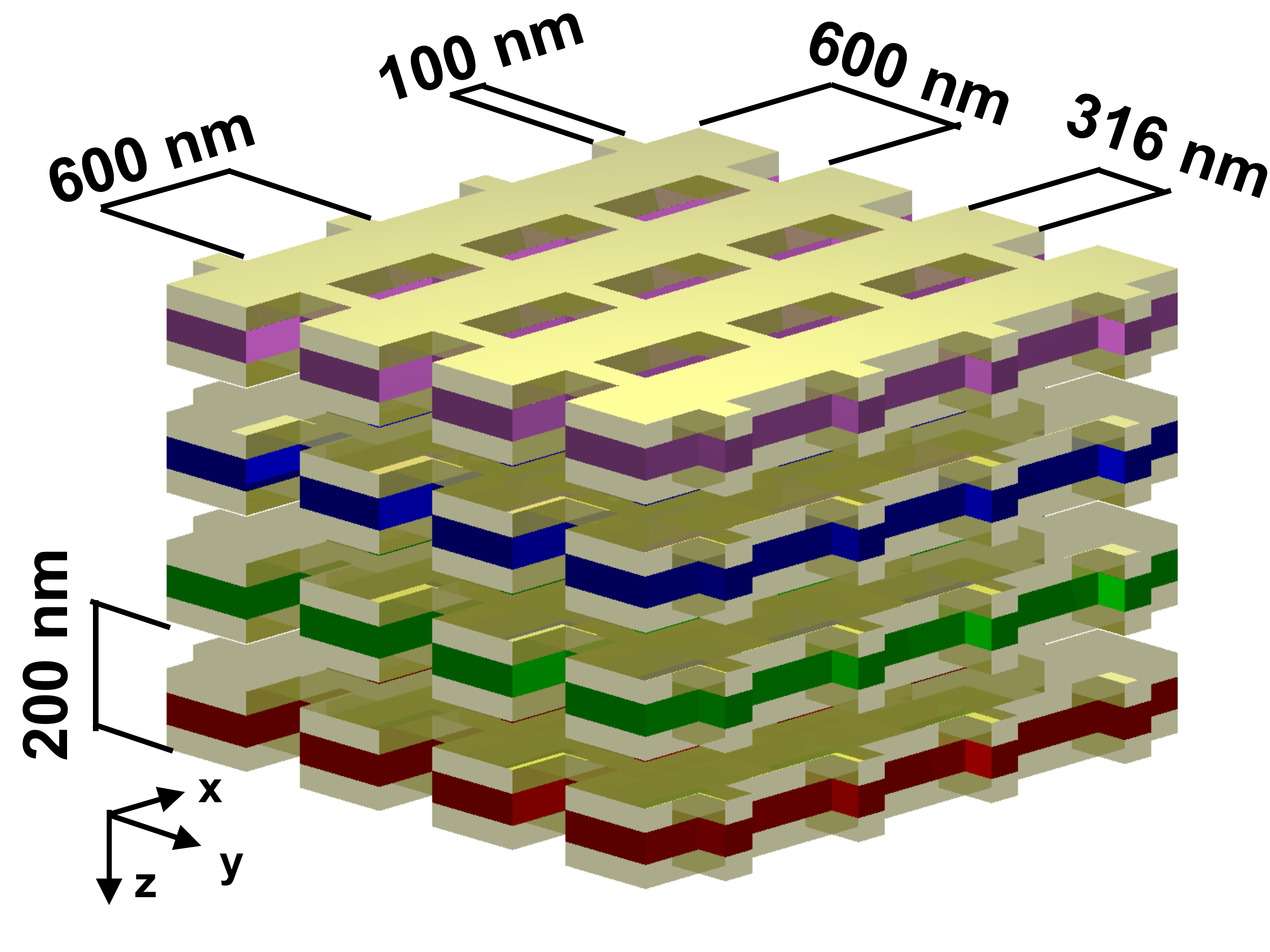}
\caption{Artistic view of the supercell metamaterial. The
geometrical parameters explicitly stated were kept constant
throughout the manuscript. Parameters subject to modifications are
the thicknesses of the metallic layers and the thicknesses and the
permittivity of the dielectric spacers. The latter are indicated by
colors and serve to identify the different functional layers in the
super cell. The principal propagation direction is $z$, the incident
field is TE ($y$) polarized with no variation in $y$-direction.}
\label{Geometry}%
\end{figure}

These additional degrees of freedom are introduced here by
considering metamaterials composed of supercells. The relevant
geometry is shown in Fig.~\ref{Geometry}. A supercell is formed by
four functional layers in $z$ direction. Each functional layer
consists of two metallic layers separated by a dielectric spacer and
is periodic in the transverse $x$ and $y$ directions. The functional
layers are distinguished by slightly different layer thicknesses and
dielectric constants of the spacer. Each functional layer exhibits
the fishnet structure where the geometrical data is taken from
literature\cite{fishnet}. The fishnet is promising since it allows
observing left-handed waves at reasonable low dissipation
\cite{fishnetopti}. Furthermore, it can be fabricated by using
existing nanotechnology as a bulk material consisting of many
functional layers \cite{staple}. This technology permits for a
variation of the optical properties of subsequent layers \cite{bulk}
as required in our approach. In what follows we optimize the
additional free parameters (thicknesses, dielectric constant of
spacer) such that the metamaterial exhibits the predefined optical
properties. We present examples for two relevant cases. The first
concerns the design of a metamaterial exhibiting a dispersion
relation with a circular iso-frequency curve being equivalent to an
effective medium with $ n_{\textrm{eff}} (k_x ,k_y ) \approx  - 1$
and the second a metamaterial where light propagates
diffractionless.

\begin{figure}[h]
\includegraphics[width=12cm]{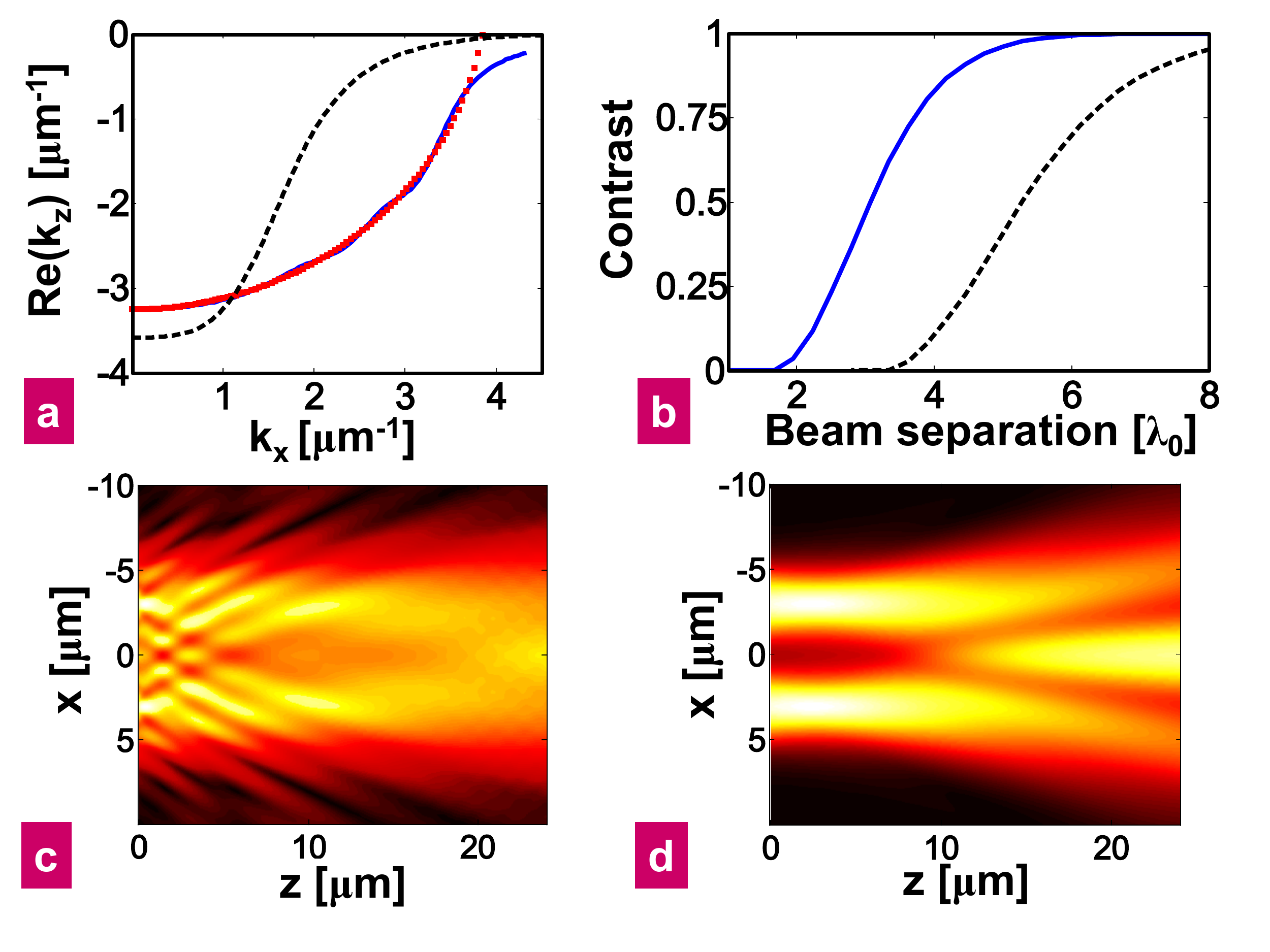}
\caption{(a) Real part of the longitudinal wavevector component
$k_z$ as a function of the transverse wave vector component $k_x$
(iso-frequency contour) for the original fishnet (black dashed line)
and the optimized supercell (blue solid line). The best circular fit
is superimposed (red dotted line). (b) The emerging contrast in the
object plane upon imaging of two Gaussian beams as function of their
initial separation  (original fishnet - black dashed line; super
cell - blue solid line). (c) Field distribution behind the exit
facet of a $2.4~\mu\text{m}$ thick fishnet MM slab (three
supercells) if two closely spaced Gaussian beams are placed with
their waist in the object plane ($-1.2~\mu\text{m}$) and for a
separation of $6~\mu\text{m}$; (d) same scenario but with an
optimized super cell .}
\label{Figure_1}%
\end{figure}
For referential purpose Fig.~\ref{Figure_1}(a) shows the real part
of $k_z$ as a function of  $k_x$ at a fixed wavelength
($\lambda=1.44\mu$m) for the original fishnet structure where all
functional layers are identical\cite{fishnet}. All simulations are
performed using a plane wave expansion technique (see Appendix AI).
Material properties for silver were taken from literature
\cite{JohnsonChristy}. The incident field is TE polarized along the
$y$ direction; being the proper polarization for the fishnet to show
the desired resonant behavior. The fishnet is neither expected to
operate for TM polarization nor for wave fields spatially confined
in $y$ direction. The thicknesses of the metal and the dielectric
layers were $h=45~\text{nm}$ and $h=30~\text{nm}$, respectively. The
permittivity of the dielectric spacer was $\epsilon=1.90$.
Subsequent functional layers are separated by air; an assumption
which is not essential and is lifted below. From
Fig.~\ref{Figure_1}(a) it is evident that the medium is anisotropic
but neither uni- nor biaxial because spatial dispersion strongly
affects light propagation. Most detrimental for the imaging purpose
is the change in sign of the curvature of the iso-frequency contour
at $k_x=1.6~\mu\text{m}^{-1}$. At this inflection point diffraction
changes from anomalous to normal preventing reasonable imaging since
a lens must compensate for the normal diffraction in free space by a
suitable anomalous diffraction. This is proven by imaging two finite
Gaussian beams where the image contrast quickly deteriorates with
smaller separation (see Fig.~\ref{Figure_1}(b)). Details on the
calculation can be found in Appendix AIV. The appearance of normal
diffraction for higher spatial frequencies as well as the angular
dependent absorption limits the resolution of the metamaterial slab
to only $4\lambda_0$.

Now by applying numerical routines (see Appendix AII) we optimized
the thicknesses of the metal and the dielectric layers in each
functional layer individually. The first target function was a
circular iso-frequency contour for the real parts of the wavevector;
comparable to that of an isotropic medium with $n= -1$. All
geometrical parameters of the optimized structure are documented in
Appendix III. The resulting iso-frequency curve along with a best
circular fit is shown in Fig.~\ref{Figure_1}(a). The effective index
of the medium corresponding to the radius of the best fit amounts to
$n_{\textrm{eff}}=-0.92$. We note that although the medium acts
optically isotropic, the geometry of the unit cell is not. Optical
isotropy is achieved by globally balancing the spatial dispersion in
each functional layer such that all together it effectively
disappears. The contrast of two Gaussian beams imaged by this
optimized metamaterial slab is shown in Fig.~\ref{Figure_1}(b).
Resolution is almost doubled, though it is not yet sub-wavelength.
Predominant reason for this is the yet remaining angular dependent
absorption. However, this restriction can be lifted by imposing
further constraints; as shown below. The superiority of the
supercell metamaterial over the conventional is shown in
Figs.~\ref{Figure_1}(c) and (d), where the field distributions
behind a 2.4 $\mu$m slab are shown if two closely spaced Gaussian
beams are imaged. The waist width \mbox{($1/e$~-~descent)} of the
Gaussian beams was $4\lambda_0 / (2\pi) = 0.91\mu\text{m}$ each and
their separation amounted to $6\mu\text{m}$. The waists are located
$1.2~\mu\text{m}$ in front of the slab. As can be seen, for the
ordinary fishnet the spots are not resolved whereas the supercell
fishnet allows for their clear imaging in the anticipated focal
position $1.2~\mu\text{m}$ behind the slab.

To verify that the constraints imposed on the optical performance
can be rather freely chosen we detail a second example. Now the
target function is a longitudinal wave vector component (real and
imaginary part of  $k_z$) that is independent of the transverse
component $k_x$ leading to diffractionless propagation. To ensure
that the optimized supercell may be technologically implemented, we
choose $\textrm{Al}_\textrm{x}\textrm{Ga}_{1-\textrm{x}}\textrm{As}$
as the dielectric spacer material \cite{GaAs}. Now subsequent
functional layers are separated by a polymer with $\epsilon=2.4$
already used in fabricating stacked metamaterials\cite{bulk}. The
thickness of the
$\textrm{Al}_\textrm{x}\textrm{Ga}_{1-\textrm{x}}\textrm{As}$ layer
was $40~\text{nm}$. Its permittivity was allowed to vary in the
range $\epsilon=8.41...11.42$. This variation is achievable by a
proper stoichiometric ratio of aluminium and gallium. Another free
parameter to be optimized was the thickness of the metal layers.
Wavelength of operation was $1.56\mu\text{m}$.
\begin{figure}[h]
\includegraphics[width=12cm]{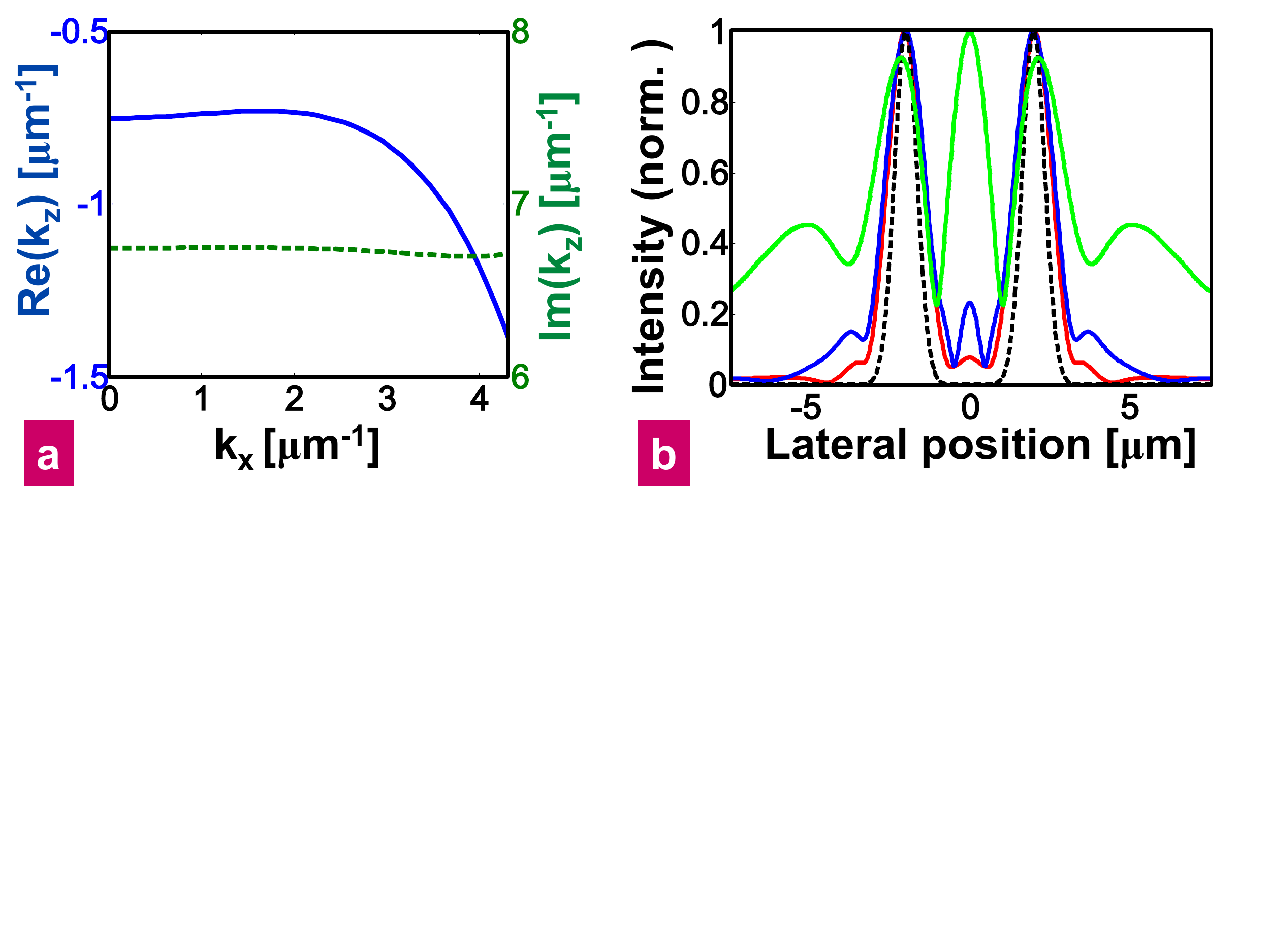}
\caption{(a) Real (blue solid line) and imaginary (green dashed
line) part of the longitudinal wavevector component $k_z$ as
function of the transverse wave vector component $k_x$
(iso-frequency contour) for the optimized super cell. Target
function of the optimization was an angle independent propagation
constant. (b) Field at the exit facet of a 2.4$\mu$m (red solid
line; three supercells) and of a 4.8$\mu$m (blue solid line; six
supercells) if two closely spaced Gaussian beams (separation is
$4\mu$m) are placed at the entrance facet (black dashed line). For
comparison the field distribution upon free space propagation over a
distance of about 4.8$\mu$m is also shown (green solid line).}
\label{Figure_2}%
\end{figure}

The optimized iso-frequency curves are shown in
Fig.~\ref{Figure_2}(a). All geometrical details are documented in
Appendix III. It can be seen, that the imaginary part is almost
constant, whereas the real part is sufficiently constant up to
$\approx$ 2.5$\mu$m$^{-1}$. If both quantities would be invariant,
each component of the plane wave spectrum of a finite object would
exhibit the same phase advance and dissipation upon traversing the
metamaterial slab, hence the field distribution at the entrance and
exit facet of the metamaterial would be identical.

A verification of almost diffractionless propagation through the
optimized metamaterial is documented in Fig.~\ref{Figure_2}(b).
There, two closely spaced Gaussian beams (width $1.0~\mu\text{m}$)
were placed at the entrance facet of the metamaterial. The field
distribution at the exit facet for a metamaterial thickness of 2.4
$\mu$m and of 4.8 $\mu$m shows only negligible distortions due to
normal diffraction for high spatial frequencies. When compared to
the field distribution upon free space propagation, it is evident
that diffraction is arrested. Details of the field evolution are
documented in Appendix V.

In conclusion, we have shown that metamaterials can be tailored to
exhibit certain optical functionalities inaccessible with natural
occurring materials. Rather than meaningless effective material
parameters the dispersion relation of Bloch waves constitutes the
quantity on which constraints are imposed. Here the optimization
procedure has been carried out to achieve either an almost perfect,
but to date no subwavelength, imaging or diffractionless propagation
through a metamaterial. We stress that different target functions
can be easily chosen. The required, optimized metamaterials can be
fabricated with standard planar technologies already at hand.
Constraints imposed on the two-dimensional iso-frequency curve in
the current work are only due to the finite computational resources;
but in principle they can be imposed on the three-dimensional
iso-frequency curve and even on the full dispersion relation. Then
one ultimate goal can be achieved, namely the propagation of a light
bullet where spreading in space and time is arrested. The ability to
mould the flow of light as revealed in this contribution is only a
prelude of a genuine class of metamaterials studies that are about
to follow.

\noindent \textbf{Acknowledgements}

 This work was partially financially supported by
the Federal Ministry of Education and Research (project Metamat),
the Deutsche Forschungsgemeinschaft (Grant No. RO 3640/1-1) and the
Thuringian State Government (project Mema).

\newpage
\noindent \textbf{Appendix}

\noindent \textit{A\textrm{I} Plane wave expansion technique}

All simulations in this work were performed by using a home-made
code that computes the dispersion relation for a unit cell
periodically arranged in the three dimensional space. It takes
explicit advantage of this periodicity by expanding all fields and
the structure into a plane wave basis. The computation consists of
two steps. At first the unit cell is reduced to a certain number of
finite layers in which the permittivity is invariant along a
predefined principal propagation direction. The eigenmodes
propagating in each layer are then solved for by a devoted
eigenvalue problem\cite{Turunen, Lifeng}. Assembling these solutions
in a scattering matrix and solving for eigensolutions of this
scattering matrix allows to find the longitudinal component of the
wave vector (propagation constant) of the Bloch eigenmodes as a
function of the frequency and  the transverse wave vector
component\cite{LiHo}. In our simulations $25\text{x}25$ plane waves
were retained in the simulation; ensuring sufficient numerical
accuracy. The advantage of this method is the simple physical
interpretation of the propagation constant as a measure for the
phase advance of an eigenmode along the propagation direction. The
number of modes one solves for is related to the number of plane
waves retained in the expansion. In the analysis, only the eigenmode
with the lowest imaginary part of the propagation constant is
considered. It will dominate the light transport since all other
modes suffer a strong absorption. All propagation constants appear
as pairs with the same magnitude but opposite sign in the real as
well as in the imaginary part; reflecting a forward and a backward
propagating wave. For preventing unphysical solutions, one has to
choose only that eigenmode which ensures an exponentially decaying
solution. Opposite sign in the real and the imaginary part signify
an opposite phase and group velocity. Reflection and transmission
from a finite slab are computed by imposing usual boundary
conditions instead of Bloch periodic boundaries on the fields in the
different spatial domains constituting the problem. They are the
incident and the transmitted region (eigenmodes are plane waves) and
the periodic region (eigenmodes are Bloch modes). Illumination of
the structure by finite beams, such as Gaussians, is modeled by
decomposing the illumination into a set of
planes waves and propagating each plane wave individually. \\

\noindent \textit{A\textrm{II} Optimization routines}

Having in mind that the spatial dispersion in each functional layer
of the super cell shall be compensated to achieve a predefined
dispersion relation, we computed at first the respective dispersion
relations for an individual functional layer depending on a single
characteristic parameter, such as the thickness of the dielectric
intermediate layer or the thickness of the metallic layers while
keeping all the other parameters constant. From these solutions a
first qualified guess for the arrangement of the functional layers
in the super cell could be derived by enforcing that the mean of the
dispersion relation matches the predefined criteria. However, this
rather simplistic approach does not provide satisfying results; but
serves only for the purpose to generate an initial set of data for
the subsequent optimization. For this purpose a nonlinear
least-squares optimization method (subspace trust-region algorithm
based on the interior-reflective Newton method\cite{Opti}) was used
to optimize the free parameters in the geometry. By relying on the
qualified guess about 15 iterations (which correspond to 135
evaluations of the structure's response) were usually required to
approach the optimized solutions. We note that slightly deviating
initial conditions might lead to different solutions for the
geometrical parameters which do allow, however, for the same
dispersion relation.

\noindent \textit{A\textrm{III} Detailed structural parameters}

Here we provide detailed information on the precise structural
parameters of the optimized metamaterials yielding i) an isotropic
(imaging)
 or ii) flat iso-frequency curve (diffractionless propagation). In both
cases each supercell consists of four functional layers of
$0.2~\mu\text{m}$ constant thickness with slightly different
structural parameters given in the tables below. The entire
metamaterial slab consists of three supercells such that the overall
slab thickness amounts to $2.4~\mu\text{m}$. The metallic layers
consist of silver, whereby the dielectric spacer (the layer in
between the two metallic layers) is $\text{MgF}_\text{2}$ with
$\epsilon=1.90$ in case i) and it is
$\text{Al}_\text{x}\text{Ga}_\text{1-x}\text{As}$ in case ii). The
permittivity of $\text{Al}_\text{x}\text{Ga}_\text{1-x}\text{As}$
was taken from literature and varies in the range $\epsilon = 8.41
\ldots 11.42$ according to the chosen stoichiometric ratio x. The
remaining space within the unit cell is only suggested to be
completed by a polymer ($\epsilon = 2.40$) in case ii).

\begin{table}[h]
\begin{tabular}{c|c|c|c} \label{tab1}
        & ~thickness~ & ~thickness of~ & ~permittivity of~ \\
~\#layer~ & ~of metal~ &  ~dielectric spacer~ & ~dielectric spacer~
\\ \hline 1 & 75.0~nm & 29.7~nm & 1.90 \\
2 & 72.4~nm & 27.9~nm & 1.90 \\
3 & 74.1~nm & 26.0~nm & 1.90 \\
4 & 76.9~nm & 24.2~nm & 1.90 \\
\end{tabular} \caption{Structural parameters of the optimized
metamaterial which exhibits a spherical angular dispersion
relation.}
\end{table}

\begin{table}[h]
\begin{tabular}{c|c|c|c} \label{tab2}
        & ~thickness~ & ~thickness of~ & ~permittivity of~ \\
~\#layer~ & ~of metal~ &  ~dielectric spacer~ & ~dielectric spacer~
\\ \hline
1 & 49.8~nm & 40.0~nm & 9.21 \\
2 & 52.1~nm & 40.0~nm & 9.70 \\
3 & 33.4~nm & 40.0~nm & 11.05 \\
4 & 34.3~nm & 40.0~nm & 11.06 \\
\end{tabular} \caption{Structural parameters of the optimized
metamaterial which exhibits diffractionless propagation.}
\end{table}

\noindent \textit{A\textrm{IV} Imaging properties}

The imaging properties of a metamaterial slab were evaluated
by illuminating the structure by two spatially separated Gaussian
beams. To calculate the transmitted field behind the slab, the
illuminating field was decomposed into its plane wave spectrum and each
plane wave was propagated through the structure as described in
Appendix AI section. The waist position was located
$1.2~\mu\text{m}$ in front of the metamaterial slab. This position
exactly corresponds to the front-side focus of a perfect lens with
$n = -1$ and a thickness of $2.4~\mu\text{m}$. These dimensions
 match nearly that of the optimized metamaterial which consists of
three supercells (12 functional layers of thickness
$0.2~\mu\text{m}$ each) and which exhibits an effective refractive
index of $n_\text{eff}=-0.92$ (see main text body for reference).

The field distribution of the electrical field at the waist position
is given by \begin{equation} \label{sm1} E_{y}(x; z=-1.2\mu\text{m})
= \exp\left[ -\dfrac{(x - S_x)^2}{\sigma^2} \right] + \exp\left[
-\dfrac{(x + S_x)^2}{\sigma^2} \right]
\end{equation} with the waist width $2\sigma$ and the beam
separation $2S_x$. To quantify the resolution limit of a
metamaterial lens the beam separation $2S_x$ was continuously
decreased whereby the waist width was kept constant with a
reasonable value. Now, tracing the field in the focus position
$(x_f, z_f)$ allows to evaluate the imaging contrast by
\begin{equation} \label{sm2} C = \dfrac{|E_y(x_f, z_f)|^2 - |E_y(0,
z_f)|^2}{|E_y(x_f, z_f)|^2 + |E_y(0, z_f)|^2}. \end{equation} Of
course, in general there exist two foci at $(x_{f}, z_f)$ and
$(-x_{f}, z_f)$, but because of the symmetry of the illuminating
field they are identical. The waist width of the Gaussian beams was
chosen to be $4\lambda_0 / (2\pi) = 0.91~\mu\text{m}$ in case of the
optimized structure. Further decreasing this size only negligibly
affects the transmitted field, since the remaining angular dependent
absorption of the metamaterial acts as a low pass filter and
therefore it limits the smallest achievable feature size in
transmission. Evaluating the imaging contrast of the original
fishnet structure the waist width of the Gaussian beams was set to
$12\lambda_0 / (2\pi) = 2.75~\mu\text{m}$. Otherwise the introduced
beam distortions in the transmitted field become to large and a
clear determination of the backside foci becomes impossible.
\\
\newpage
\noindent \textit{A\textrm{V} Diffractionless propagation}

\begin{figure}[h]
\includegraphics[width=13cm]{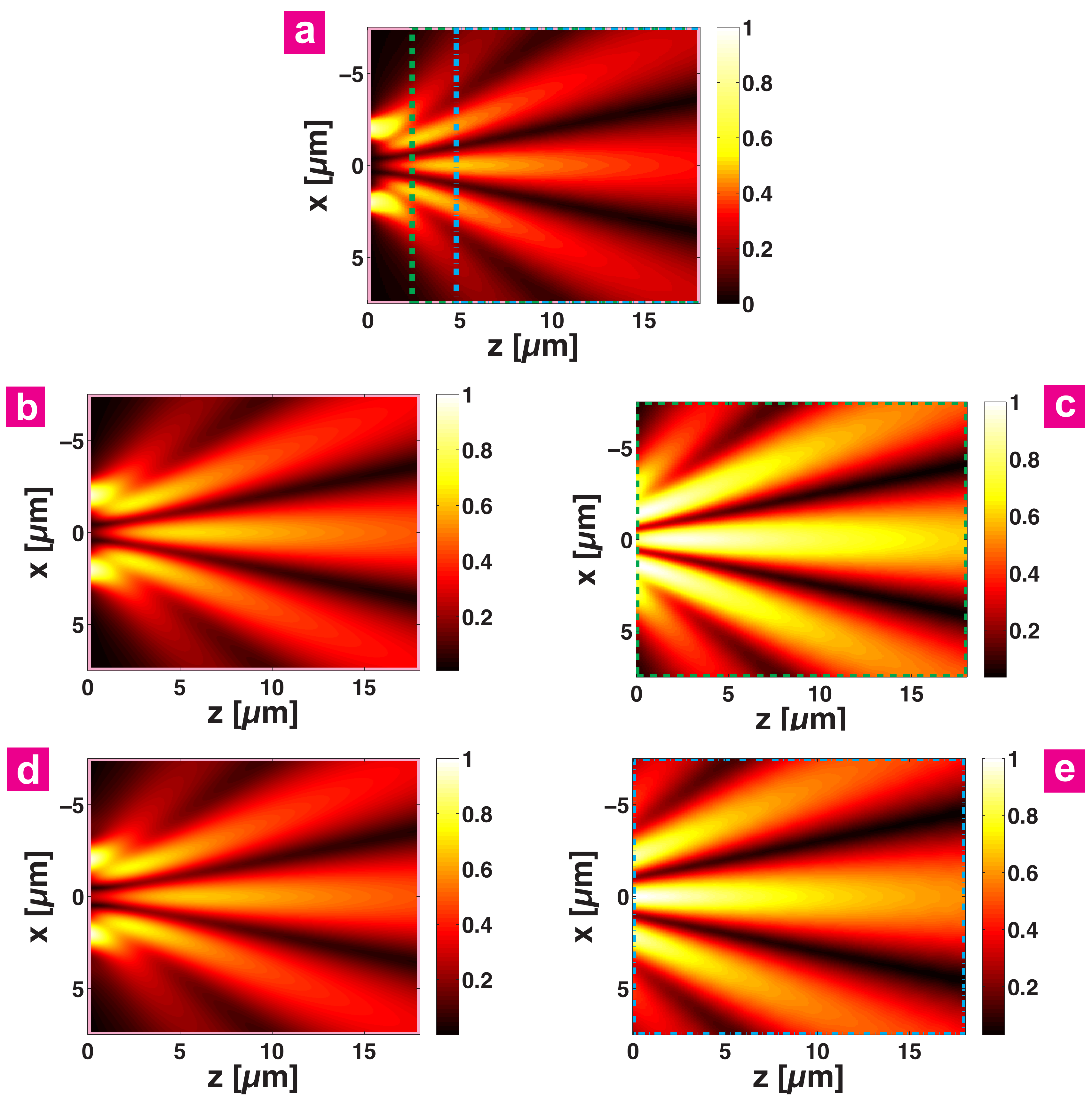}
\caption{(a) The normalized electrical field amplitude
 of a double peak Gaussian source (waist width $4\lambda_0 /
(2\pi) = 1.0~\mu\text{m}$) during propagation in air at a wavelength
$1.56~\mu\text{m}$  for referential purpose. The beam
separation is $4~\mu\text{m}$. The waist is located at
$z=0$. By inserting the optimized metamaterial slab structure of
thickness (b) $2.4~\mu\text{m}$ and (d) $4.8~\mu\text{m}$
diffraction is suppressed and the propagating field distribution
directly behind the exit facet ($z=0$) changes only negligibly when
compared to figure (a). The field plots in figure (c) and (e) show
for comparison the case, when the metamaterial slab is removed and
the field propagates the same distance of $2.4~\mu\text{m}$ and
$4.8~\mu\text{m}$ in free space, respectively. The colored boxes
indicate sections of associated field distributions.}
\label{Figure_4}%
\end{figure}

\end{document}